\documentclass[referee]{aa} 
\begin{document}

   \thesaurus{12     
              (12.03.4;  
               12.04.1;  
               12.05.1;  
               12.07.1;  
               12.12.1)} 
   \title{The fluid mechanics of dark matter formation}

   \subtitle{Why does Jeans's (1902 \& 1929) theory fail?}

   \author{Carl H. Gibson}

   \offprints{C. H. Gibson}

   \institute{Departments of Applied Mechanics and 
Engineering Sciences and\\ Scripps Institution of 
Oceanography, University of California
at San Diego,\\ La Jolla, CA 92093-0411, USA\\
              email: cgibson@ucsd.edu
             }

   \date{Received July 1, 1998;  }

   \maketitle

   \begin{abstract}

Jeans's (1902 \& 1929) gravitational instability criterion 
gives truly spectacular errors in its predictions of cosmological
structure formation according to Gibson's (1996) new theory.  
It is suggested that the linear perturbation stability analysis
leading to the acoustic Jeans length scale criterion
$L \ge L_J \equiv V_s /(\rho G)^{1/2}$ is quite irrelevant to gravitational
condensation, which is intrinsically nonlinear and nonacoustic.
Instead, condensation is limited by viscous or turbulent 
forces, or by diffusivity,
at Schwarz length scales $L_{SV}$, $L_{ST}$, or $L_{SD}$, respectively,
whichever is larger, independent of $L_{J}$ and the sound speed $V_s$.
By these new criteria, cosmological structure formation begins
in the plasma epoch soon after matter dominates energy, at 
$L \approx L_{SV} \equiv (\gamma \nu / \rho G)^{1/2}$ scales
corresponding to protosuperclusters, decreasing to protogalaxies
at the plasma-gas transition, where $\gamma$ is the rate-of-strain
of the expanding universe, $\nu$ is the kinematic viscosity, $\rho$ is
the density, and $G$ is Newton's gravitational constant.
Condensation of the primordial gas occurs at mass scales a trillion
times less than the Jeans mass to form a `fog' of 
micro-brown-dwarf (MBD) 
particles that persist as the galactic baryonic dark matter, as
reported by Schild (1996) from quasar-microlensing studies.  
Nonbaryonic dark matter condensation is prevented by its enormous
diffusivity at scales smaller than $L_{SD} \equiv (D^2 /\rho G)^{1/2}$,
where $D$ is the diffusivity, so it forms outer halos of galaxies,
cluster-halos of galaxy clusters, and supercluster-halos.

   \end{abstract}

%

\section{Introduction}
Jeans's theory fails because it is linear. Linear theories
typically fail when applied to nonlinear processes; for example,
applications of the linearized Navier-Stokes equations
to problems of fluid mechanics give laminar solutions,
 but observations show that actual flows
are turbulent when the Reynolds number is large.

Cosmology (e.g.; \cite{pbl93}, \cite{pad93}) relies exclusively 
on Jeans's theory in modelling the
formation of structure by gravitational instability. Because
$L_J$ for baryonic (ordinary) matter
 in the hot plasma epoch following the Big Bang is larger
than the Hubble scale of causality $L_H \equiv ct$, where
$c$ is the speed of light and $t$ is the time, no baryonic structures
can form until the cooling plasma forms neutral gas.  
Star and galaxy formation models invented to accommodate 
both Jeans's theory and the
observations of early structure formation have employed a variety
of innovative maneuvers and concepts.  Nonbaryonic ``cold
dark matter'' was invented to permit nonbaryonic condensations in
the plasma epoch with gravitational potential wells that
could guide the early formation of baryonic galaxy masses.  
Fragmentation theories were proposed to produce 
$M_{\sun}$-stars rather
than Jeans-superstars at the
$10^{5} M_{\sun}$ proto-globular-cluster Jeans mass
of the hot primordial gas.

Both of these concepts have severe fluid mechanical difficulties
according to the \cite{gib96} theory.  Cold dark matter cannot
condense at the galactic scales needed 
because its nonbaryonic, virtually collisionless, nature
requires it to have an enormous diffusivity 
with supergalactic $L_{SD}$ length scales.
Fragmentation theories (e.g., \cite{low76}) are based on a faulty
condensation premise that 
implies  large velocities and a powerful turbulence regime
that would produce a first generation of large stars with 
minimum mass
determined by the turbulent Schwarz scale $L_{ST} \equiv
\varepsilon ^{1/2} /(\rho G)^{3/4}$, 
where $\varepsilon$ is the viscous
dissipation rate of the turbulence, with a flurry of starbursts,
supernovas, and metal production that is not observed
in globular star clusters.  The population of small, long-lived,
metal-free, globular cluster stars observed is strong evidence
of a quiet, weakly turbulent formation regime.

\section{Jeans's acoustic theory}
Jeans considered the problem of gravitational
condensation in a large body of nearly constant
density, nearly motionless gas.  Viscosity and diffusivity
were ignored.  The density and momentum conservation
equations were linearized by  dropping second order
terms after substituting mean plus fluctuating values
for the density, pressure, gravitational potential, and velocity.
Details of the derivation are given in 
many cosmological texts (e.g.; \cite{kol94}, p342) so
they need not be repeated here.
The mean gravitational force
$\nabla \phi$ is assumed to be
zero, violating the Poisson equation
\begin{equation}
\nabla ^2 \phi = 4 \pi G \rho ,
\end{equation}
where $\phi$ is the gravitational potential, in
what is known as the Jeans swindle.
Cross-differentiating the linearized perturbation
equations produces a single, second order differential
equation satisfied by Fourier modes propagating at the
speed of sound $V_s$.  From the dispersion equation
\begin{equation}
\omega^2 = V_s^2 k^2 - 4 \pi G \rho ,
\end{equation}
where $\omega$ is the frequency and $k$ is the wavenumber,
a critical wavenumber $k_J = (4 \pi G \rho / V_s^2)^{1/2}$
exists, called the $\emph{Jeans wavenumber}$.  For $k$
less than $k_J$, $\omega$ is imaginary and the mode
grows exponentially with time.  For $k$
larger than $k_J$, the mode is a propagating sound wave.
Density was assumed to be a function only of pressure
(the barotropic assumption).  

Either the barotropic assumption or the linearization of
the momentum and density equations are sufficient to reduce
the problem to one of acoustics.  Physically,
sound waves provide density nuclei at wavecrests
that can trigger gravitational condensation if their time
of propagation $\lambda/V_s$ for wavelength
$\lambda$ is longer than the \emph{gravitational free fall
time} $\tau_g \equiv (\rho G)^{-1/2}$.  Setting the two times
equal gives the Jeans gravitational instability
criterion: gravitational condensation occurs only
for $\lambda \ge L_J$.

Jeans's analysis fails to account for the effects of gravity,
diffusivity, or fluid mechanical forces
upon nonacoustic density maxima and density minima; that is, points
surrounded on all sides by either lower or higher density.
These move 
approximately with the
fluid velocity, not $V_s$,  (\cite{gib68}).  The evolution of such 
\emph{zero gradient points}
and associated \emph{minimal gradient surfaces} is critical
to turbulent mixing theory (\cite{gak88}).  Turbulence
scrambles passive scalar fields such as temperature, chemical species
concentration and density to produce nonacoustic extrema,
saddle points, doublets, saddle lines and minimal
gradient surfaces.  A quasi-equilibrium
develops between convection and diffusion 
at such zero gradient points and minimal gradient
surfaces that is the basis of a universal 
similarity theory of turbulent mixing (\cite{gib91})
analogous to the universal similarity theory of Kolmogorov
for turbulence.
Just as turbulent velocity fields are damped by viscosity
at the Kolmogorov length scale $L_K \equiv (\nu/\gamma)^{1/2}$,
where $\nu$ is the kinematic viscosity and $\gamma$ is the
rate-of-strain, scalar fields like temperature are damped
by diffusivity at the Batchelor length scale $L_B \equiv
(D/\gamma)^{1/2}$, where $D$ is the molecular diffusivity.  This
prediction has been confirmed by laboratory experiments
and numerical simulations (\cite{gak88}) for 
the range $10^{-2} \le Pr \le 10^5$, where the Prandtl
number $Pr \equiv \nu/D$. 

On cosmological length scales, density fields scrambled
by turbulence are not necessarily dynamically passive but 
may respond to gravitational forces.  In the density
conservation equation
\begin{equation}
\partial \rho/\partial t + v_i (\partial \rho/\partial x_i)
= D_\mathrm{eff}\partial ^2 \rho/\partial x_j \partial x_j
\end{equation}
the effective diffusivity of density 
$D_\mathrm{eff} \equiv D - L^2 /\tau_g$
 is affected by gravitation in
the vicinity of minimal density gradient features, and 
reverses its sign to negative if the feature size $L$ is larger than
the diffusive Schwarz scale $L_{SD}$ (\cite{gib98}).
$L_{SD} \equiv (D^2/\rho G)^{1/4}$ is derived
by setting the diffusive velocity $v_D \approx D/L$
 of an isodensity surface
a distance $L$ from a minimal gradient configuration
 equal to the gravitational velocity
$v_g \approx L/\tau_g$.  Thus, nonacoustic density maxima in 
a quiescent, 
otherwise homogeneous, fluid
are absolutely unstable to gravitational condensation,
and nonacoustic density minima are absolutely unstable to
void formation, on scales larger than $L_{SD}$.

Jeans believed from his analysis (\cite{jns29}) that
sound waves with $\lambda \ge L_J$ would grow in
amplitude indefinitely, producing unlimited kinetic energy
from his gravitational instability.  This is clearly
incorrect, since any wavecrest that collects a finite
quantity of mass from the ambient fluid will also collect
its zero momentum and become a nonacoustic density nucleus.
From the enormous Jeans mass values indicated at high temperature,
he believed he had proved his speculation that the cores of galaxies
consisted of hot gas (emerging from other Universes!) and not stars,
which could only form in the cooler (smaller $L_J$) 
spiral arms, thrown into cold outer space
by centrifugal forces of the spinning core.  The
concepts of \emph{pressure support} and \emph{thermal support}
often used to justify Jeans's theory are good examples
of bad dimensional analysis, lacking any proper physical basis.

\section{Fluid mechanical theory}
Gravitational condensation on a nonacoustic density
maximum is limited by either diffusion
or by viscous, magnetic or turbulent forces at diffusive,
viscous, magnetic, or turbulent Schwarz scales $L_{SX}$,
whichever is largest, where $X$ is $D,V,M,T$,
respectively (\cite{gib96, gib98}).  
Magnetic forces are assumed to be
unimportant for the cosmological conditions of interest.
Gravitational forces $F_g \approx \rho^2 G L^4$ equal
viscous forces $F_V \approx \rho \nu \gamma L^2$ at
$L_{SV} \equiv (\nu \gamma/\rho G)^{1/2}$, and turbulent
forces $F_T \approx \rho (\varepsilon)^{2/3}L^{8/3}$ at
$L_{ST} \equiv \varepsilon ^{1/2} /(\rho G)^{3/4}$.
Kolmogorov's theory is used to estimate the turbulent
forces as a function of length scale $L$.  

The criterion (Gibson 1996, 1997a, 1997b; 
Gibson and Schild 1998a, 1998b) 
for gravitational condensation or void
formation at scale $L$ is therefore
\begin{equation}
L \ge \left( L_{SX} \right)_{\mathrm {max}};
X = D,V,M,T . 
\end{equation}

\section{Structures in the plasma epoch}
Without the Jeans constraint, structure formation 
begins in the early stages of the hot plasma
epoch after the Big Bang
when decreasing viscous forces first permit gravitational
decelerations and sufficient time has elapsed for the
information about density variations to propagate; that
is, the decreasing viscous
Schwarz scale $L_{SV}$ becomes smaller than the
increasing Hubble scale $L_H \equiv ct$, where $c$ is the
velocity of light.  Low levels of temperature fluctuations
of the primordial gas indicated by the COsmic microwave
Background Experiment (COBE) satellite 
($\delta T/T \approx 10^{-5}$) constrain the velocity
fluctuations $\delta v /c \ll 10^{-5}$ to 
levels of very weak turbulence.  Setting the
observed mass of superclusters $\approx 10^{46}$ kg 
equal to the Hubble mass $\rho L_H^3$
computed from Einstein's equations (\cite{win72}, 
Table 15.4) indicates the time of first structure
was $\approx 10^{12}$ s, or $30\,000$ y (\cite{gib97b}).

Setting $L_H \approx 3 \, 10^{20}$ m (10 kpc) = $L_{SV}$
gives $\nu \approx 6 \, 10^{27}$ $ \mathrm{m^2 \, s^{-1}}$
with $\rho \approx 10^{-15}$ $ \mathrm{kg \, m^{-3}}$ and
$\gamma \approx 1/t = 10^{-12}$ $ \mathrm{rad \, s^{-1}}$.
Such a large viscosity suggests a neutrino-electron-proton
coupling mechanism, presumably
through the Mikheyev-Smirnov-Wolfenstein (MSW)
effect (\cite{bak97}), supporting the Neutrino-98 claim that neutrinos
have mass. 

The viscous condensation mass $\rho L_{SV}^3$ decreases
to about $10^{42}$ kg (\cite{gib96}) as the Universe expands
and cools to the plasma-gas transition at $t \approx 10^{13}$
s, or $300\,000$ y, based on Einstein's equations 
to determine $T$ and $\rho$ and assuming
the usual dependence of viscosity $\nu$ on temperature $T$
(\cite{win72}).  Assuming gravitational decelerations
that are possible always occur, we see that protosupercluster,
protocluster, and protogalaxy structure formation 
should be well
underway before the emergence of the primordial gas.

\section{Primordial fog formation}
The first condensation scales of the primordial
gas mixture of hydrogen and helium are the maximum size
Schwarz scale, and an initial length scale 
$L_{IC} \equiv (RT/\rho G)^{1/2}$
equal to the Jeans scale $L_J$ but independent
of Jeans's linear perturbation stability analysis,
and acoustics, where $R$ is
the gas constant of the mixture.
From the ideal gas law $p/\rho = RT$ we see that
density increases can be compensated by pressure
increases with no change in temperature in a uniform
temperature gas, and that gravitational forces
$F_g \approx  \rho ^2 G L^4$
will dominate the resulting pressure gradient forces
$F_p \approx p L^2 = \rho R T L^2$ for length scales
$L \ge L_{IC}$.  Taking $R \approx 5\,000$ 
$\mathrm{m^2 \, s^{-2} \, K^{-1}}$, 
$\rho \approx 10^{-18}$ $\mathrm{kg \, m^{-3}}$ m  
(\cite{win72})
and $T \approx 3\,000$ K gives a condensation
mass $\rho L_{IC}^3 \approx 10^5 M_{\sun}$, the
mass of a globular cluster of stars.  Because
the temperature of the primordial gas was observed
to be quite uniform by COBE, we can expect the
protogalaxy masses of primordial gas emerging from the plasma
epoch to immediately fragment into proto-globular-cluster
(PGC) gas objects on $L_{IC} \approx 3 \, 10^{17}$ m (10 pc) 
scales, with subfragments at 
$\left( L_{SX} \right)_{\mathrm {max}}$.

The kinematic viscosity $\nu$ of the primordial gas
mixture decreased by a factor of about a trillion from
plasma values at transition, to $\nu \approx 2.4 \, 10^{12}$
$\mathrm{m^2 \, s^{-1}}$ assuming the density within
the PGC objects are about $10^{-17}$ $\rm kg \> m^{-3}$.
Therefore, the viscous Schwarz scale
$L_{SV} \approx (2.4 \, 10^{12} \, 10^{-13}
/ 10^{-17} \, 6.7 \, 10^{-11})^{1/2} = 1.9 \, 10^{13}$ m,
so the viscous Schwarz mass $M_{SV} \approx 
L_{SV}^3 \rho = 6.8 \, 10^{22}$ kg, or
$M_{SV} = 6.8 \, 10^{24}$ kg using $\rho = 10^{-18}$.
The turbulent Schwarz mass $M_{ST} \approx 8.8 \, 10^{22}$ kg
assuming $10 \%$ of the COBE temperature fluctuations are due to
turbulent red shifts ($[(\delta v / c ) / ( \delta T/T)] = 10^{-1}$)
as a best estimate. 

	We see that the entire universe of primordial H-He gas turned
to fog soon after the plasma-gas transition, with primordial
fog particle (PFP) mass values in the range $10^{23}$ to $10^{25}$
kg depending on the estimated density and turbulence levels of the gas.
The time required to form a PFP is set by the time required
for void regions to grow from minimum density points and maximum
density saddle points to surround and isolate the condensing
PFP objects (\cite{gib98}).  Voids grow as rarefaction waves with
a limiting maximum velocity $V_s$ set by the second
law of thermodynamics, so the minimum PFP formation time
is $\tau_{PFP} \le (L_{SX})_{\mathrm{max}}/V_s$, or about $10^3$ y.
Full condensation of the PFP to form a dense core near hydrostatic
equilibrium requires a much longer time, near the gravitational
free fall time $\tau_g \approx 2 \, 10^6$ y.  

Radiation heat transport during the PFP condensation  period
before the creation of dust should have permitted cooling to
temperatures near those of the expanding universe.  After about
a billion years hydrogen dew point and freezing point temperatures
(20-13 K) would be reached, forming the micro-brown-dwarf conditions
expected for these widely separated ($10^3-10^4$ AU) 
small planetary objects that comprise
 most of the baryonic dark matter of the present
universe and the materials of construction
for the stars and heavy elements. 
 Because such frozen objects occupy an angle
of less than a micro-arcsecond viewed from their average separation 
distance, they are invisible to most observations except by
gravitational microlensing, or if a nearby hot star brings
these volatile comets out of cold storage.

\section{Observations}
A variety of observations confirm the new theory that
fluid mechanical forces and diffusion
limit gravitational condensation (\cite{gib96}), and confute
 Jeans's (1902 \& 1929) acoustic criterion:
\begin{itemize}
      \item quasar microlensing at micro-brown-dwarf frequencies
(\cite{sch96}),
      \item tomography of dense galaxy clusters
indicating diffuse (nonbaryonic) superhalo dark matter at $L_{SD}$ scales
with $D_{nb} \approx 10^{28} \mathrm{m^2 \, s^{-1}}$ (\cite{tyf95}),
\item the Gunn-Peterson missing gas sequestered as PFPs,
\item the dissipation of `gas clouds' in the $Ly-\alpha$ forest, 
\item extreme scattering events, cometary globules, FLIERS, 
ansae, Herbig-Haro `chunks', etc.
   \end{itemize} 

Evidence that the dark matter of galaxies is
dominated by small planetary mass objects has been
accumulating from reports of many observers
 that the multiple
images of lensed quasars twinkle at corresponding high frequencies.
After several years spent resolving a controversy
about the time delay between images Q0957+561A,B,
to permit correction for any
intrinsic fluctuations of the light intensity
of the source by subtraction of the
properly phased images, \cite{sch96} announced that
the lensing galaxy mass comprises 
 $\approx 10^{-6} M_{\sun}$ ``rogue planets'' that are 
``likely to be the missing mass.''  

Star-microlensing studies from the Large Magellanic Cloud
have failed to detect lensing at small planetary mass frequencies,
thus excluding this quasar-microlensing
population as the  Galaxy halo missing mass 
(\cite{alc98}, \cite{ren98}).  However, the exclusion
is based on the unlikely assumption that the number density of such
small objects is uniform.  The population must have mostly
primordial gas composition since no cosmological model predicts
this much baryonic mass of any other material, and
must be primordial since it constitutes the material of
construction, and an important stage, in the condensation of the gas to form
stars.  Gravitational aggregation
is a nonlinear, self-similar, cascade process 
likely to produce an extremely
intermittent lognormal spatial distribution 
of the PFP number density,
 with mode value orders of magnitude smaller than
the mean.  Since star-microlensing from a 
small solid angle produces
a small number of independent samples, the observations
estimate the mode rather than the mean, resolving the observational
conflict (\cite{gs98}).

\section{Conclusions}
   \begin{enumerate}
      \item Jeans's gravitational instability criterion
$L \ge L_J$ is irrelevant
to gravitational structure formation in cosmology
and astrophysics, and is egregiously misleading
in all of its applications.
      \item The correct criterion for gravitational structure
formation is that $L$ must be larger than the
largest Schwarz scale; that is, 
$L \ge \left( L_{SX} \right)_{\mathrm {max}}$,
 where  $X$ is $D,V,M,T$, depending on whether
diffusion or viscous, magnetic or turbulent forces
limit the gravitational effects.
      \item Structure formation began in the plasma
epoch with protosupercluster to
protogalaxy decelerations.
\item Gravitational condensations began soon after
the plasma-gas transition, forming micro-brown-dwarfs,
clustered in PGCs,
 that persist as the dominant
dark matter component of inner galactic halos (50 kpc).
\item   The present fluid mechanical theory and its 
cosmological consequences
regarding the forms of baryonic and nonbaryonic 
dark matter  (\cite{gib96}) is well supported
by observations, especially the quasar-microlensing
of \cite{sch96} and his inference that the lens galaxy mass
of Q0957+561A,B is dominated by small rogue planets
(interpreted here as PFPs).
\item  Star-microlensing studies that rule out
MBDs as the Galaxy missing mass 
(\cite{alc98}, \cite{ren98}), contrary to the
quasar-microlensing evidence and the present theory, are
subject to extreme undersampling errors from
 their unwarranted assumption of a uniform number density
 distribution, rather than extremely intermittent
lognormal distributions expected from 
nonlinear aggregational cascades of such small objects as
they form nested clusters and stars (\cite{gs98}).
   \end{enumerate}

\begin{acknowledgements}
      Numerous helpful suggestions were provided by Rudy Schild.
\end{acknowledgements}

\end{document}